\documentclass[%
twocolumn,longbibliography,preprintnumbers,
nofootinbib,
 amsmath,amssymb,
 aps,
prb,
]{revtex4-2}
\usepackage{graphicx}
\usepackage{amsmath}
\usepackage{amssymb}
\usepackage{ulem}
\usepackage{xcolor}

\begin{document}
\title{Global population crisis scenarios predicted by a general nonlinear dynamical model}


\author{Alessio Zaccone}
\affiliation{Department of Physics ``A. Pontremoli'', University of Milan, via Celoria 16, 20133 Milan, Italy.}

\author{Kostya Trachenko}
\altaffiliation{Deceased.}
\affiliation{School of Physical and Chemical Sciences, Queen Mary University of London, Mile End, London, U.K.}

\begin{abstract}
We show that a simple nonlinear differential equation (originally studied in the physics of disordered systems) is able to mathematically describe the global population growth over the past 12000 years. Different regimes of population growth since the early Neolithic until today are shown to be all solutions to the same nonlinear differential equation in its various limits. These also include the well-known Malthus (exponential) and Verhulst (logistic) growth regimes, as well as von Foerster's ``doomsday'' formula. All these limits correspond to neglecting higher-order terms in a more general nonlinear dynamic model described by the proposed nonlinear differential equation. 
While the older models may provide valid fittings to limited time intervals in the global population growth curve in time, their clearly approximate nature prevents them from being predictive over longer periods of time. The proposed comprehensive solution of the proposed model is instead well suited to provide predictions for future scenarios. These include a scenario where the global population could halve as early as 2064 under a deliberately conservative, worst-case assumption that carrying-capacity constraints become abruptly active today.

\end{abstract}

\maketitle
The first modern attempt to mathematically describe the time evolution of biological populations goes back to Malthus, who identified two central facts which regulate the growth of a population based on the number of its individual elements. The first is that each new element is generated by a sexual or asexual process (birth), and the second is that each element ceases to be a member of the population (death). Of course there are many complex factors which determine the population-averaged birth and death rates, but assuming a closed population (with no inward or outward fluxes) nothing forbids one to define an average birth rate per element per unit time and an average death rate per element per unit time. This leads to the following equation for the rate of change of the number of elements in the population, $y$, originally proposed by Malthus in 1798 \cite{Malthus}:
\begin{equation}
    \frac{dy}{dt}= b\,y - d\,y = r\, y \label{Malthus}
\end{equation}
where $b$ and $d$ are the birth rate per element per unit time and the death rate per element per unit time, respectively, and $r = b-d$ is known as the productivity of the individual element. 
If both $b$ and $d$ are constants, so is also $r$, and the solution to Eq. \eqref{Malthus} is an exponential growth if $b>d$ or an exponential decrease if $b < d$. In demography, the exponential growth (Malthusian) scenario has often been used to motivate discussions of socioeconomic drivers of fertility rates and long-term sustainability in order to avert a potential doomsday due to overpopulation outrunning food supply (the latter growing arithmetically and not exponentially). The Malthusian or neo-Malthusian perspective has remained popular until our time, with many prominent modern thinkers, including e.g. C. Lévi-Strauss \cite{Levi}, adhering to its world view.
However, short after Malthus' influential essay, it became clear that assuming a constant $r$ is too strong an assumption, because a population will have a natural tendency to self-regulate its growth based on available resources. This led Verhulst \cite{Verhulst} to hypothesise that $r$ is a decreasing function of $y$, and in particular to assume a linear decrease: $r(y)= r_0(1-y/\kappa)$, where $r_0$ is a constant and $\kappa$ is the carrying capacity of the environment, or the maximum population that the environment can sustain. With this correction in Eq.\eqref{Malthus}, the unbounded exponential growth is replaced by a limited growth, which ends with a plateau or stable equilibrium, $y \rightarrow const$ for $t \rightarrow \infty$. This is the famous logistic growth curve with its typical sigmoidal shape, which has been used many times to fit empirical data on animal populations in ecology or bacterial colonies in petri dishes \cite{Lotka,Murray}.  

Taking a different view, von Foerster and co-workers argued \cite{Foerster} that intelligent beings would organize in a strongly linked society, thus reducing the original problem to that of a two-person game between the population and nature as its opponent. Based on this game-theoretical assumption, the growth of the population would lead to more effective ways to counter hazards and to produce resources. As a result, the productivity $r$ of the population would be an increasing function of the population size $y$, for example a power-law increasing function of $y$. von Foerster and co-workers applied this idea to fit the then available data of global population growth, and found an excellent fitting provided by a hyperbolic function: $y(t)=A/(D-t)$, where $D=2026$ is the "doomsday" year at which the population size was predicted to diverge to infinity. This hyperbolic form is the exact solution to the modified differential equation upon setting $r(y)= \alpha y$ in Eq. \eqref{Malthus}, with $\alpha$ a constant (see also \cite{Yakovenko}).
It was however recognized already in the following decades that the global population trend was diverting from the von Foerster's formula prediction, thus averting the predicted doomsday scenario in 2026. 

In addition to classical growth laws, a large literature addresses population crises via scenario-specific extensions of logistic-type models, including formulations with explicit delays, thresholds, or multiple regulatory functionals; see e.g. Perevaryukha \cite{Perevaryukha2021} and references therein. In contrast, the objective of the present work is to show that several empirically observed growth regimes can arise within a single analytically tractable nonlinear rate-feedback equation, without introducing explicit time delays or piecewise regulatory mechanisms.

A recent important work \cite{Sojecka2024} has shown that the global population evolution as a function of time over the past 12000 years has gone through various growth regimes, including short intervals of decline e.g. corresponding to the Black Death plague epidemics. Excluding these very short decline periods, the most prominent regimes, as visible in Fig. 1, are described by the following functions: (i) simple exponential growth (Malthus), corresponding to straight lines in the semi-logarithmic plot; (ii) flat or horizontal lines $y(t) = const$, corresponding to Verhulst or logistic plateaus; (iii) stretched-exponential function (SEF), $\exp\!\left[(t/\tau)^\beta\right]
$ with $\beta <1$ and (iv) compressed-exponential function (CEF),  $\exp\!\left[(t/\tau)^\beta\right]
$ with $\beta > 1$. All these best-fitting functions, including CEF and SEF, are particular solutions to the Eq. \eqref{TZ} proposed in this paper.

\begin{figure}[ht]
    \centering
    \includegraphics[width=\linewidth]{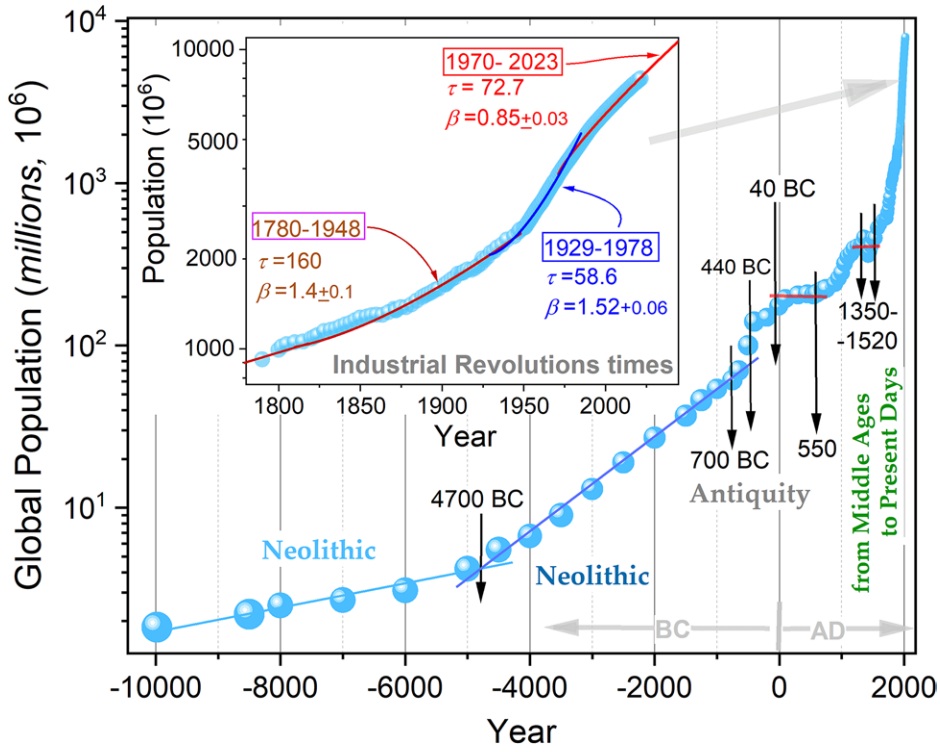}
    \caption{Evolution of the global population over the past 12000 years until now. Symbols refer to empirical measurements and continuous lines are the best fits obtained with the functional forms (i)--(iv) discussed in the text. Simple exponential growth is given by straight lines, while Verhulst logistic plateaus appear as flat horizontal lines. The inset shows the stretched-exponential (SEF) and compressed-exponential (CEF) growth regimes.
Reproduced from Ref.~\cite{Sojecka2024} under the Creative Commons Attribution 4.0 International License (CC BY 4.0). No changes were made to the original figure.}
    \label{fig1}
\end{figure}

A similar stretched-exponential function function, with only a difference of sign inside the argument of the exponential,   $y(t) \sim \exp(-t/\tau)^\beta$, has been studied for almost two centuries in physics, at least since when Kohlrausch 
studied the electric discharge as a function of time of a prototypical capacitor, the Leyden jar \cite{Kohlrausch,Phillips}. 
Since then, the stretched-exponential time decay, also referred to as the Kohlrausch function, has been observed in countless experiments and simulations of time-dependent phenomena in glasses and other amorphous solids \cite{Ngai,kob_book,zaccone2023}, such as dielectric relaxation \cite{Loidl,Bingyu}, mechanical relaxation experiments \cite{Donth,Wang}, and in the time-dependence of atomic-scale cooperative motions \cite{Ruta}. More recently, also the compressed exponential decay in time, with $\beta >1$, has been experimentally detected in the atomic dynamics of metallic glasses \cite{Ruta,Kob} as well as in colloidal gels \cite{McKenna}.

A theoretical model, which is able to produce all the above functional trends (i)-(iv) as its solutions, represents the productivity rate with the following function of the population size $y$: $r(y)=\exp(K y)/\tau$, where $K$ is a real number, which could be positive or negative, and $\tau$ is a time constant. In what follows we will use $y$ to denote a normalized population within each regime (defined explicitly below), which removes dimensional ambiguity in the rate function.
The corresponding differential equation thus reads as (compare to Eq. \eqref{Malthus}):
\begin{equation}
    \frac{dy}{dt}= b\,y - d\,y = r(y)\, y= \frac{y}{\tau}\exp(K y). \label{TZ}
\end{equation}
Here and in the following, the population variable is normalized within each growth regime as
\begin{equation}
y \equiv \frac{N(t)}{N(t_0)},
\end{equation}
where $t_0$ is the initial time of the considered regime. Hence $y$ is dimensionless and the argument $Ky$ in the exponential is dimensionless as well (with $K$ dimensionless).

This provides a highly general nonlinear model of population growth, because it allows the productivity rate $r(y)$ to either increase with the population size $y$ for $K > 0$ or to decrease with the population size for $K < 0$. 
In particular, SEF solutions to Eq. \eqref{TZ} are obtained for $K<0$, while CEF solutions are obtained for $K>0$. 

In the physics and chemistry of glasses and supercooled liquids, Eq. \eqref{TZ} (sometimes referred to as the Trachenko-Zaccone (TZ) equation) has an extra minus sign: $ \frac{dy}{dt}= - \frac{y}{\tau}\exp(K y)$ \cite{Trachenko_2021,Ginzburg} because it describes relaxation, i.e. a delay or lag in the response of a material to an external (mechanical, electrical or magnetic) perturbation. Hence, the physically measurable response is described by a decaying function of time, the simplest being the simple exponential (Debye) decay, and, more common in liquids and glasses, the Kohlrausch relaxation function already quoted above \cite{Ngai,Phillips,Zaccone_2020}.
The key physical mechanism that gives this equation in the context of glassy dynamics, is that local events (correlated atomic motions) do not occur independently but influence the rate of future events (for liquids, glasses and other systems, this influence is due to long-ranged elastic stresses in the material). In glasses, $K<0$ and $K>0$ in the TZ equation correspond to local relaxation events slowing down and speeding up the relaxation of future events, respectively. In Ref. \cite{Ginzburg}, it has been shown numerically that a 1:1 relation exists between the Kohlrausch exponent $\beta$ in the relaxation functions $y(t) \sim \exp(-t/\tau)^\beta$ (note the minus sign) and the TZ parameter $K$, for both SEF $\beta < 1$ and CEF $\beta > 1$.  The same 1:1 correspondence relations exists for the  parameter $K$ in Eq. \eqref{TZ} and the $\beta < 1$ SEF and $\beta > 1$ CEF growth functions $\exp\!\left[(t/\tau)^\beta\right]
$ (now mind the plus sign inside the exponential). In particular, in Eq. \eqref{TZ}, $K<0$ corresponds to $\beta < 1$ SEF growth in time, while $K>0$ corresponds to $\beta > 1$ CEF growth in time.

Equation \eqref{TZ} can be readily shown to recover the Malthus equation, Eq. \eqref{Malthus}, upon linearizing the productivity rate about $K y=0$, $\exp(K y) \approx 1$ setting $r = 1/\tau = const$. 
Upon linearizing it for $K<0$, we obtain $\exp(K y) \approx (1 - s y)$, where $s \equiv -K$ is a positive real number. In this limit, Eq. \eqref{TZ} recovers the Verhulst logistic growth.
If, instead, we linearize the exponential for $K>0$, $\exp(K y) \approx (1 + b y)$, where $b>0$, Eq. \eqref{TZ} reduces to the Riccati equation with constant coefficients $\frac{dy}{dt}= a\,y^2 + b \,y$, where $a \equiv K/\tau$ and $b \equiv 1/\tau$, which is solved by the von Foerster's doomsday formula \cite{Foerster}, $y(t)=A/(D-t)$ with $A = 1 + b c_0$ and $D= (\frac{1}{a} - bc_0 -1)$ in the limit of $1 + b c_0 \gg b t$, where $c_0$ is the integration constant. Hence, in this limit, Eq. \eqref{TZ} predicts a doomsday scenario with a divergence of the global population to infinity at a time $y =(\frac{1}{a} - bc_0 -1) $.

In its general form, and without linearizing the exponential, Eq. \eqref{TZ} admits exact solutions in terms of the exponential integral function $\textrm{Ei}(x)$ and its inverse \cite{Abramowitz,Lebedev}, as follows:
\begin{equation}
y(t) = -\frac{1}{K} \textrm{Ei}^{-1}\left(\textrm{Ei}(-K) +t\right)
\end{equation}
where we used a dimensionless time $t \equiv t/\tau$.
For values of the argument $|x| < \frac{\mu}{\ln(\mu)}$, the inverse exponential integral function can be written as a series expansion:
\begin{equation}
\textrm{Ei}^{-1}(x) = \sum_{n=0}^\infty \frac{x^n}{n!} \frac{P_n(\ln(\mu))}{\mu^n}
\end{equation}
where $\mu = 1.4513692...$ is the Ramanujan-Soldner constant and $P_n$ are polynomials defined recursively by $P_0(x) = x,\ P_{n+1}(x) = x(P_n'(x) - nP_n(x))$. We have checked that the solution produced in this way closely matches the numerical solution (apart from very few discontinuous points which are due to the complexity of numerically inverting the exponential integral function).

As already mentioned above, Eq. \eqref{TZ} among its solutions, admits both the SEF and the CEF growth regimes, alongside simple exponential, von Foerster, and sigmoidal when the exponential is linearized. Therefore, Eq.~\eqref{TZ} provides a unified nonlinear population-growth equation able to reproduce multiple empirically observed regimes as exact or controlled limiting solutions. We are not aware of any other differential equation or dynamical model which can achieve such a comprehensive mathematical description. Also, by making $K$ a function of time, Eq. \eqref{TZ} is able to mathematically describe the crossovers between different growth regimes, e.g. from SEF to CEF and viceversa, and of course also from Malthusian to Verhulst and viceversa, and, in fact, between any of the regimes (i)-(iv) in Fig. \ref{fig1}, from and to each other. For glassy relaxation this was demonstrated in Ref. \cite{Trachenko_2021}.

In the above model, we considered $r(y)=\exp(K y)$ as the most generic function describing the productivity rate of global population as the result of complex social, economic and environmental factors. The value of $K$ is the only parameter which modulates the growth and is a lumped function of all these complex factors. Using the physical meaning of $K$ in condensed matter systems, we can say that the sign of $K$ is set by whether the occurrence of local events slow down or speed up the occurrence of future events \cite{Trachenko_2021}.

\begin{figure}
    \centering
\includegraphics[width=0.9\linewidth]{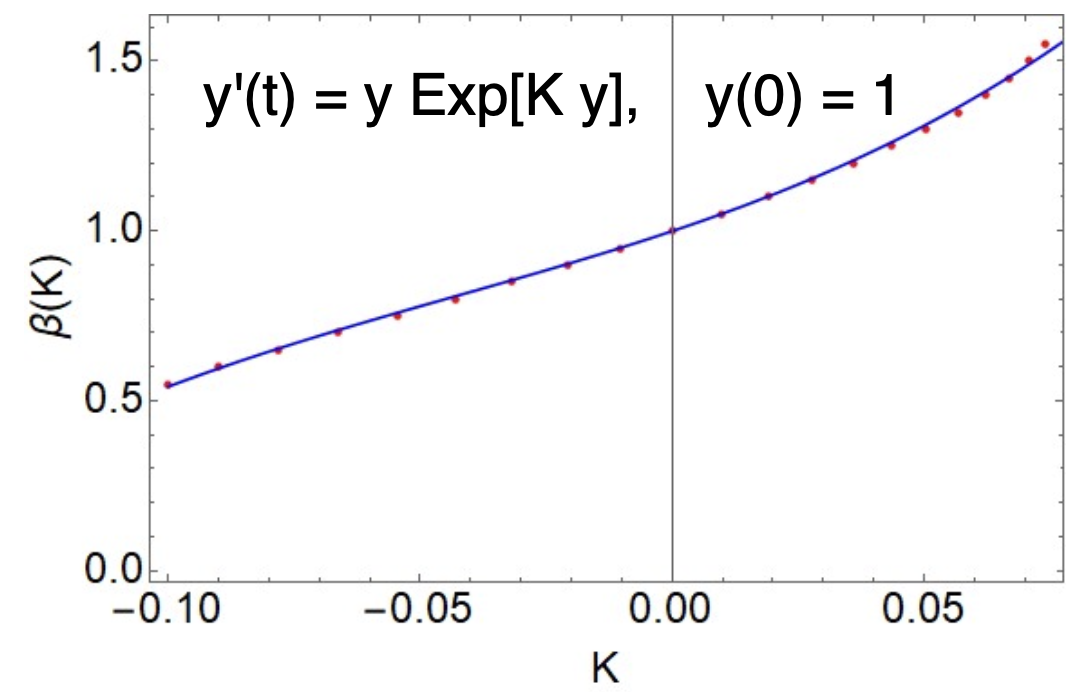}
    \caption{Mapping between the CEF/SEF parameter $\beta$ and the TZ equation parameter $K$ obtained by numerical error minimization. Symbols correspond to numerical minimization results, while the continuous line is a guide to the eye.}
    \label{mapping}
\end{figure}

Based on the numerical correspondence between the SEF/CEF exponent $\beta$ and the parameter $K$ in Eq. \eqref{TZ}, reported in Fig. \ref{mapping}, the crossover around 1970 from CEF to SEF growth corresponds to a crossover from $K\simeq 0.072$ to $K=-0.032$ in Eq.~\eqref{TZ}.

\begin{figure*}[ht]
    \centering
    \includegraphics[width=6.6in]{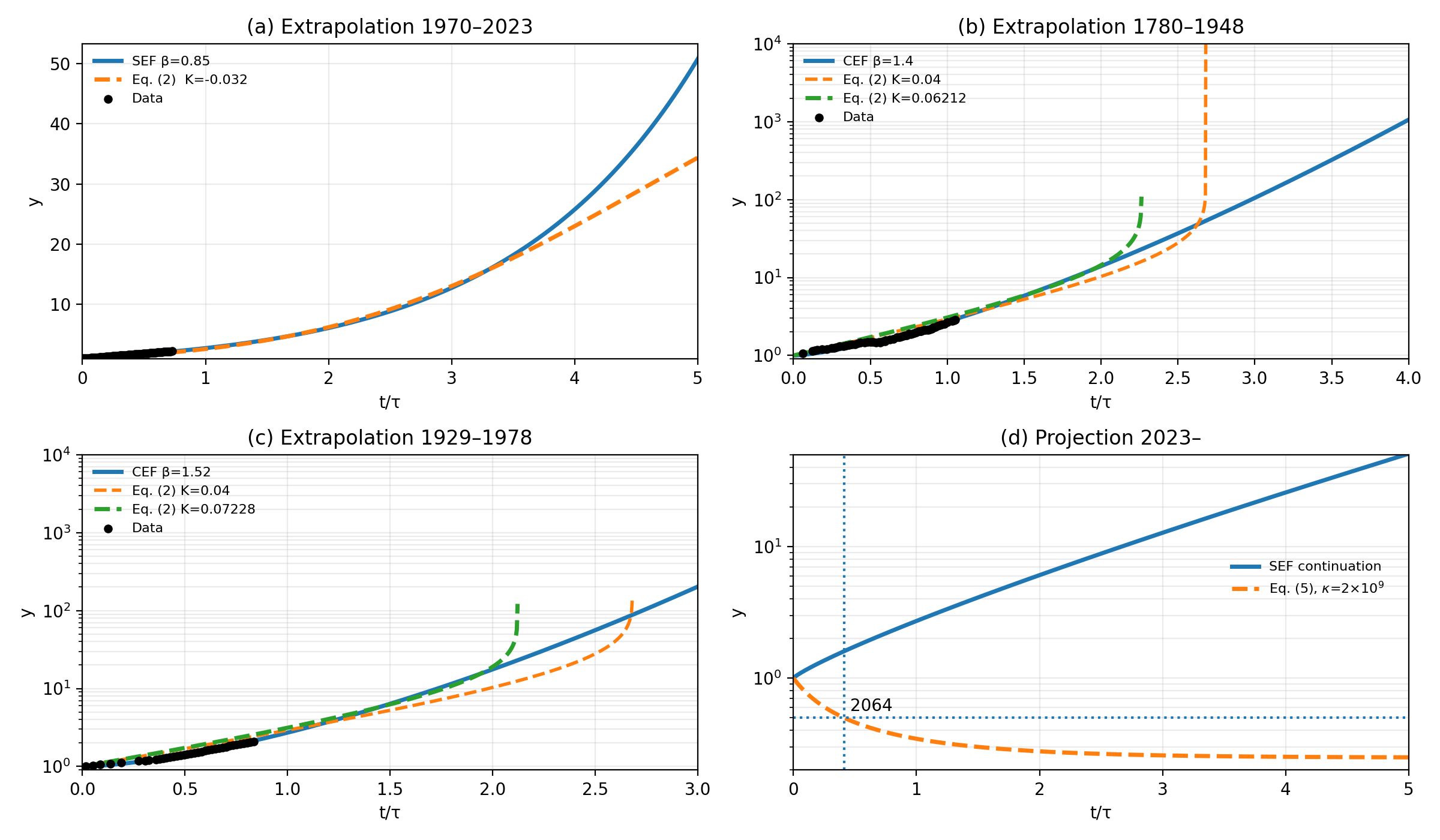}
    \caption{Extrapolations of global population growth regimes. The markers are empirical population data points from Ref.~\cite{Sojecka2024} (Appendix table therein), normalized in the same way as the curves: $y\equiv N(t)/N(t_0)$ and $t\equiv (T-t_0)/\tau$, where $T$ is calendar year and $\tau$ is the characteristic time used in the corresponding SEF/CEF fits \cite{Sojecka2024}. Thus $t=0$ in each panel corresponds to the start year $t_0$ of the considered regime.
In panels (a)--(c), solid curves are the best-fit SEF/CEF functions reported in Ref.~\cite{Sojecka2024}, while dashed curves are numerical solutions of Eq.~\eqref{TZ}. Panel (a) (1970--2023): SEF with $\beta=0.85$ and $\tau=72.7$ yr; Eq.~\eqref{TZ} with $K=-0.032$. Panel (b) (1780--1948): CEF with $\beta=1.40$ and $\tau=160$ yr; Eq.~\eqref{TZ} with $K=0.04$ (illustrative) and $K=0.062117398$ (best fit over $0<t/\tau<2$). Panel (c) (1929--1978): CEF with $\beta=1.52$ and $\tau=58.6$ yr; Eq.~\eqref{TZ} with $K=0.04$ (illustrative) and $K=0.07227915$ (best fit over $0<t/\tau<2$). Panel (d): projection from 2023, where the solid curve continues the current SEF trend (no carrying-capacity constraint) and the dashed curve shows the prediction of Eq.~\eqref{TZ_cr} if carrying-capacity constraints were to become abruptly active today. The value $\kappa=2\times 10^9$ is chosen as a deliberately conservative, worst-case estimate to illustrate the qualitative effect of such a constraint.}
    \label{fig2}
\end{figure*}

In Fig. \ref{fig2}(a), we show the comparison between the empirical best fitting SEF of the global population growth data for the period 1970-2023 (cf. Fig. \ref{fig1}) with the prediction of Eq.~\eqref{TZ} for $K=-0.032$.  The agreement is very good and the maximum deviation is within $4$\% in relative error with respect to the SEF fit. A systematic deviation appears at about 290 years from the starting point ($t/\tau=0$) set at 1970. While the prediction of Eq. \eqref{TZ} seems to slow down compared to the SEF, the predicted global population at this point (i.e. in 2260) is already 20 times the global population in 1970. This means a global population of 73.9 billions, which is clearly an unsustainable number. 

A different scenario is explored in Fig.\ref{fig2}(b) and (c), where the empirical CEF growth regimes ranging from the industrial revolution to the Second World War (CEF1), and from the 1929 crisis to 1978 (CEF2), respectively, are analyzed. Also in this case, the numerical solution to Eq. \eqref{TZ} is in fair agreement with the best fitting CEF functions, for both time periods, with best-fitting values of the parameter $K$ equal to $K=0.062117398$ (panel (b)) and $K=0.07227915$ (panel (c)), respectively. Because now we have $K>0$, as required to describe CEF, we notice the presence of a blow-up catastrophe in the numerical solution to Eq. \eqref{TZ}, at a finite time $t$. 

Extrapolating into the future, while the CEF remains bounded (even though astronomically large population sizes are reached), the solution to Eq. \eqref{TZ} displays a doomsday criticality, i.e. a sudden blow-up explosion in the population size at a finite time, as a result of the strongly nonlinear dynamics underlying Eq. \eqref{TZ}, in the regime where $r(y)$ is an exponentially increasing function of $y$. Recall that, already for $r(y) \sim y$, one gets a similar divergence in the solution to the corresponding Riccati equation as shown by von Foerster et al. \cite{Foerster} -- hence, with $r(y) \sim \exp(Ky)$ and $K>0$ this singularity is also expected to occur. According to the predictions of Eq. \eqref{TZ}, this doomsday scenario may occur as early as 2053.
However, as already mentioned, this CEF growth regime has crossed over into SEF already around 1970 (possibly thanks to worldwide increase in average education \cite{Sojecka2024,Korotayev,Impicciatore2020}), which should avert the doomsday criticality, unless the global population growth reverts to a CEF-type growth regime with $K>0$ in Eq. \eqref{TZ}.
Based on these predictions, the present study clearly confirms the possibility of a doomsday criticality of the von Foerster type, which has to be avoided by keeping the global population growth far from the $K>0$ regime in Eq. \eqref{TZ}. This could be facilitated by socioeconomic drivers known to reduce fertility rates, such as education, healthcare access, and women’s empowerment, while keeping the focus here on the mathematical implications of the growth regimes.

The orange curves in panels (b) and (c) are the best fitting values found by least-square minimization over the interval $0<t/\tau<2$, while the  blue curves with $K=0.04$ are for illustrative purposes.

To quantify goodness-of-fit while keeping the analysis minimal and consistent with the scope of this work, we evaluate residuals on a logarithmic scale,
\begin{equation}
\varepsilon_i = \ln y_i - \ln \hat y(t_i),
\end{equation}
where $y_i$ are the empirical points and $\hat y(t_i)$ are the corresponding model values. We report the log-RMSE, log-MAE and the coefficient of determination $R^2$ computed on $\ln y$. For the time windows shown in Fig.~\ref{fig2}, we obtain:
(i) panel (a) 1970--2023: log-RMSE $=0.088$, log-MAE $=0.083$, $R^2=0.854$;
(ii) panel (b) 1780--1948: log-RMSE $=0.165$, log-MAE $=0.150$, $R^2=0.594$;
(iii) panel (c) 1929--1978: log-RMSE $=0.181$, log-MAE $=0.173$, $R^2=0.254$.

While the SEF scenario with $K<0$, which best describes the population growth from 1970 until now, does not present any doomsday criticality, it is still possible that future developments may lead to a diversion from this trend. This deviation may be caused by a sudden crisis such as a global conflict, a catastrophic acceleration of the climate crisis, or a major global epidemic. As a result of such crisis, the current high efficiency in the exploitation of natural resources may significantly, and suddenly, decline. While the Earth's limited carrying capacity ($\kappa$) has not been influencing the global population growth at least since the industrial revolution, thanks to great technological and scientific developments, it cannot be excluded that the situation may change as a result of a global crisis which would compromise the efficiency of resource exploitation and management. 
After having shown the ability of Eq. \eqref{TZ} to mathematically describe all the regimes in the global population evolution over the past 12000 years, we can now explore its predictions in a hypothetical scenario where a global crisis would lead the population dynamics to depend on the Earth's carrying capacity $\kappa$. 
This hypothesis would thus lead to the following dynamic equation, where the effect of Earth's carrying capacity is made explicit \cite{Murray}:
\begin{equation}
    \frac{dy}{dt}= r(y)\, y= \frac{y}{\tau}\exp(K y)\left[1 - \frac{y}{\kappa} \right]. \label{TZ_cr}
\end{equation}
where $\kappa$ is the Earth's carrying capacity.
If the sudden crisis leading to Eq. \eqref{TZ_cr} were to set in today, we would have $K = -0.032$, since this provides the best fitting of the global population data (Fig. \ref{fig1}) from 1970 till now. By conservatively setting $\kappa = 2 \times 10^9$ according to the available estimates \cite{Lianos2016,Dasgupta,Tamburino}, the above Eq. \eqref{TZ_cr} predicts the trend shown as a dashed line in Fig. \ref{fig2}(d).
According to this prediction, there will be a halving of the current global population by 2064. Given that the current population on Earth is a bit more than 8 billions, this prediction would imply a catastrophic decline of the global population with the disappearance of about 4 billions individuals over a period of just 29 years. This scenario is approximately compatible with a global poopulation peak around 2030 recently predicted by Yakovenko \cite{Yakovenko}.

In conclusion, a single ordinary differential equation, Eq. \eqref{TZ}, is able to mathematically describe all the various regimes encountered in the global population recorded as a function of time, over the past 12000 years until now. Regimes of simple exponential growth (Malthus), logistic (Verhulst) plateaus as well as stretched-exponential and compressed-exponential growth regimes are all reliably described by Eq. \eqref{TZ} in its various limits.
Based on Eq.~\eqref{TZ}, which provides a unified nonlinear dynamical framework for multiple growth regimes, future scenarios have been analyzed.
While the current global population growth trend does not lead to a doomsday criticality, the analysis highlights the dynamical importance of remaining in a regime effectively corresponding to $K<0$ in Eq.~\eqref{TZ}. Failing to do so would lead to a doomsday criticality of the von Foerster type as predicted by the exact solution to Eq. \eqref{TZ} in Fig. \ref{fig2}(b)-(c). 
In a different scenario, a global crisis as due to a planetary war outbreak, a sudden acceleration of global warming or a vast epidemic with a high mortality rate, could lead to the modified Eq. \eqref{TZ_cr}, where the effect of Earth's limited carrying capacity becomes explicit. Considering the current estimates of the Earth's carrying capacity, if the crisis were to happen today Eq. \eqref{TZ_cr} predicts a catastrophic decline of the world's population with a net decrease by 4 billions human lives over a relatively short period of 29 years. \\

\section*{Data Availability}
The empirical population data analysed in this work are taken from Ref.~\cite{Sojecka2024} (including its appendix table). No new data were generated.

\section*{Ethical Statement}
This article does not contain any studies with human participants or animals performed by any of the authors. All simulations/analyses were carried out using computational methods, and therefore no ethical approval was required.

\subsection*{Acknowledgments}
This work is dedicated to the memory of my friend and collaborator Prof. Kostya Trachenko who passed away shortly after the writing of this paper. Many useful discussions with Dr. Valeriy Ginzburg are gratefully acknowledged. A.Z. gratefully acknowledges funding from the European Union through Horizon Europe ERC Grant number: 101043968 ``Multimech'', from US Army Research Office through contract nr. W911NF-22-2-0256, and from the Nieders{\"a}chsische Akademie der Wissenschaften zu G{\"o}ttingen in the frame of the Gauss Professorship program. 

\bibliographystyle{apsrev4-1}

\bibliography{references}
\end{document}